\documentclass[lettersize,journal]{IEEEtran}
\usepackage{amsmath,amsfonts}
\usepackage{algorithmic}
\usepackage{algorithm}
\usepackage{array}
\usepackage[caption=false,font=normalsize,labelfont=sf,textfont=sf]{subfig}
\usepackage{textcomp}
\usepackage{stfloats}
\usepackage{url}
\usepackage{verbatim}
\usepackage{graphicx}
\usepackage{cite}
\usepackage{bm}
\usepackage[table,xcdraw]{xcolor}
\usepackage{multirow}
\usepackage{orcidlink}

\begin{document}

\title{\textbf{HYDRA}: Hybrid Data Multiplexing and Run-time Layer Configurable DNN Accelerator}

\author{Sonu Kumar\textsuperscript{$\ast$}\orcidlink{0009-0000-4008-7153}, 
Komal Gupta\orcidlink{0009-0004-8705-5000}, Gopal Raut\textsuperscript{$\dagger$}\orcidlink{0000-0002-1046-9457}, \IEEEmembership{Member, IEEE},
Mukul Lokhande \orcidlink{0009-0001-8903-5159},
and Santosh Kumar Vishvakarma$\ast$\orcidlink{0000-0003-4223-0077}, \IEEEmembership{Senior Member, IEEE}

\thanks{}
\thanks{ 
The authors thank DST for providing the DST INSPIRE PhD fellowship and SMDP-C2S, the Ministry of Electronics and Information Technology (MeitY), and the Government of India for providing the necessary tools for ASIC simulation.
The authors\textsuperscript{$\ast$} are with the Centre for Advanced Electronics, IIT Indore. Dr. Gopal Raut\textsuperscript{$\dagger$} is with the Center for Development of Advanced Computing, Bengaluru-560100, India, and all authors are associated with the NSDCS Research Group, Department of Electrical Engineering, IIT Indore, Simrol-453552, India. 
\\
\textbf{Corresponding author}: Prof. Santosh Kumar Vishvakarma, \\
\textbf{E-mail:} skvishvakarma@iiti.ac.in. }
}

\markboth{}%
{Shell \MakeLowercase{\textit{et al.}}: A Sample Article Using IEEEtran.cls for IEEE Journals}

\maketitle
\begin{abstract}
Deep neural networks (DNNs) offer plenty of challenges in executing efficient computation at edge nodes, primarily due to the huge hardware resource demands. The article proposes HYDRA, hybrid data multiplexing, and runtime layer configurable DNN accelerators to overcome the drawbacks. The work proposes a layer-multiplexed approach, which further reuses a single activation function within the execution of a single layer with improved Fused-Multiply-Accumulate (FMA). The proposed approach works in iterative mode to reuse the same hardware and execute different layers in a configurable fashion. The proposed architectures achieve over 90\% of power consumption and resource utilisation improvements of state-of-the-art works, with 35.21 TOPS\/W. The proposed architecture reduces the area-overhead (N-1) times required in bandwidth, AF and layer architecture. This work shows HYDRA architecture supports optimal DNN computations while improving performance on resource-constrained edge devices. 

\end{abstract} 


\section{Introduction}
Deep neural networks, or DNNs, have gained popularity in various fields, including science, medicine, and agriculture, thanks to advances in computer technology. DNNs are highly effective for tasks involving cognition, learning, and pattern recognition, but substantial hardware and power requirements challenge their deployment at the edge. The core components of DNNs are resource-intensive, such as the fused multiply-accumulate (FMA) units that each neuron applies weighted input sums and non-linear activation functions \cite{ref1,ref2}. Traditionally, DNN computation is performed in the cloud, which introduces latency due to transmitting results back to edge devices. Deploying DNNs directly on edge devices can reduce latency, save bandwidth, and improve data security; however, limited battery power, storage, and computing resources pose significant challenges for this approach\cite{ref26}. The increasing requirement for data processing in DNNs, composed of multiple layers of interconnected neurons, exacerbates these challenges. The extensive computation must be handled by edge devices that increase the complexity in hardware and power consumption, which makes it very challenging to integrate DNNs at the edge. The edge devices enable the DNN operations at the edge with constraints like response time requirements, privacy concerns, and data transmission costs, which are very crucial for the AI applications\cite{ref9}. The design of a 1D array architecture with fused multiply-accumulate (FMA) units followed by an activation function \cite{ref10}, fused multiply-accumulate\cite{ref26} is a promising solution to these problems. Existing methods for DNN implementation on such architectures either involve implementing all layers simultaneously, which is energy-intensive and increases area, or implementing layers individually, which conserves energy but can introduce higher latency. 

The article proposes a novel architecture of a DNN accelerator that executes neural networks in a layer fashion within resource-constrained environments at Edge AI. The key contributions of this work are:

\textbf{Layer-Multiplexed DNN accelerator:}
The proposed DNN accelerator allows configuration for any depth by defining the top design parameters. The architecture uses a single layer to compute DNN sequentially by reusing the same underlined hardware. The benchmark configuration of 64:32:32:10 is evaluated and validated on the FPGA board Virtex VC-707 and with software analysis that interprets the inference accuracy results.

\textbf{Runtime Configurable Layer Architecture:}
An area-optimized, layer multiplexed FC layer is presented, which reuses a single AF with the function of PISO that parallelism the data flow and multiplexed it. The impact of saving hardware resources by (n-1) AF in each layer of the layer reused DNN configuration without the throughput loss and performance-improved Fused-Multiply-Accumulate is analysed.

The rest of this paper is organised as follows: Section 2 gives Related Work and Motivation. Section 3 describes the proposed design for the hardware-reuse technique and layer reuse for the neural network. The experimental setup is described in Section 4. Section 5 explains the performance analysis of the proposed architecture and corresponding neural network. Section 6 concludes in brief about the above work and interprets the future directions.

\begin{figure*}[!t]
     \centering
     \includegraphics[width=180mm,height=57mm]{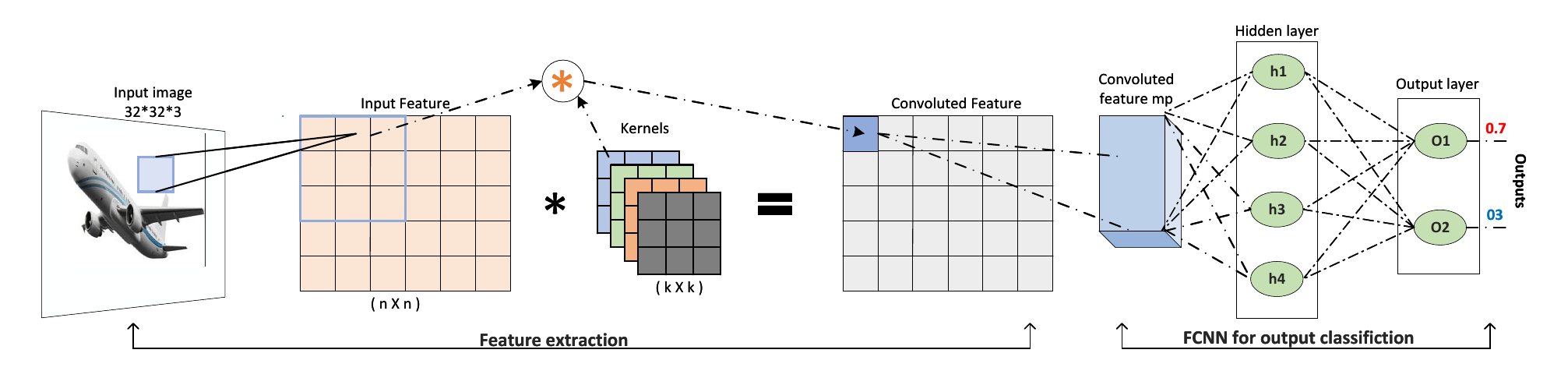}
\vspace{-5mm}
     \caption{DNN architecture with Conv. and FC. Layers. FMA computation is performed on ifmaps and Kernels.}
     \label{dnn}
\vspace{-5mm}
\end{figure*}

\begin{figure}[!t]
     \centering
     \includegraphics[width=87mm,height=100mm]{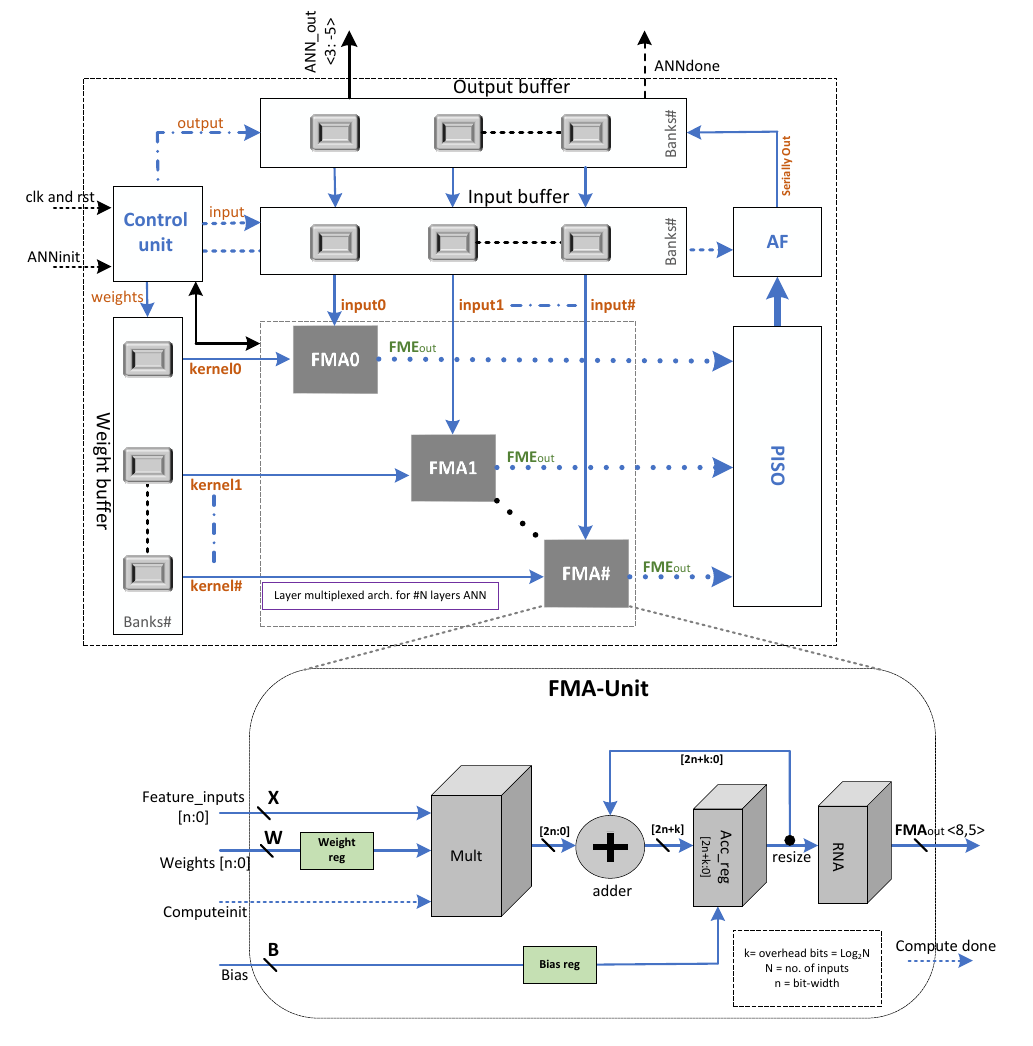}
     \caption{Runtime configurable proposed layer-reused HYDRA architecture, with subunits FMA and AF reused Layer.}
     \label{core-arch}
     \vspace{-5 mm}
\end{figure}

\section{ Related Work and Motivation}

Traditional DNN accelerators suffer from underutilising hardware underlined \cite{ref11,ref13}. Though the two-dimensional array accelerators are more fruitful for convolution and matrix multiplication, in practice, the hardware suffers from inefficient mapping techniques for ANN/RNN/fully connected layers. Thus, creating a 1-D layer based on FMA units is necessary to fit these layers more efficiently\cite{ref7,ref9}. However, it can also be noted that FMA is utilised for N*N clock cycles for N*N kernel computation, while AF is utilized only once\cite{ref16,ref17}. Hence, the infrequently used hardware can be reduced while providing an area-throughput tradeoff. Further, it could be noted that the execution of these layers is more sequential. Hence, it is more beneficial to have layer-reused architecture than fully parallel architecture, considering the resource-constrained edge environment\cite{ref15,ref20}. The proposed architecture addresses the issue by mapping different layers onto the same hardware, thus enhancing hardware efficiency and configurable architecture through data reuse. AF multiplexing further enhances the layer hardware, reducing static power consumption. Researchers have also looked at different DNN mapping methods, sparsity exploitation, and data reuse, which talk about the pros and cons of area, latency, and power \cite{ref17,ref18}. The smaller architectures can utilize fewer resources, though unsuitable considering edge devices' quick response time. Thus, instead of tiling the network over a small FMA architecture, we suggest providing at least the minimum hardware to execute a single layer simultaneously, reducing intra-dependencies in data movement. This approach further reduces the complexity of the control engine and produces a simpler, more energy-efficient architecture.

\begin{table}[!pt]
\caption{User-defined constants}
\label{user-const}
\renewcommand{\arraystretch}{1.5}
\scalebox{0.875}{
\begin{tabular}{cccc}
\hline
\textbf{Hierarchy} & \textbf{Entity} & \textbf{Pre-synthesis constants} & \multicolumn{1}{c}{\cellcolor[HTML]{FFFFFF}\textbf{Value (User-defined)}} \\\hline
1 & Layers & Layer configuration & 196:64:32:32:10 \\
2 & Max Count & \#FMA per layer & 64 \\
3 & AF & Layers (5) & 1 \\
4 & FMA & Bit-width & 8 \\
\# & Types & Integer bits & 3 (with sign bit)\\\hline
\end{tabular}}
\end{table}

\section{Proposed Work}
We designed a 1D array of 64 FMA units to implement feed-forward neural networks efficiently. The current implementation targets a 64:32:32:10 network configuration comprising four fully connected layers. The architecture includes an input layer with 64 FMA and two hidden layers (each utilising 32 FMAs) and concludes with a 10-class output layer. The accelerator will be fed 196 pixels, i.e., 14x14 (half-folded version of MNIST images.) The edge-AI hardware consists of 90\% of multiplication and addition operations, making FMA the central block. The FMA computes the product of filter maps with activation inputs and is added with bias values. A clock cycle could be saved here by preloading the bias value into the accumulator.

\subsection{Layer Reuse architecture}
Deploying deeper neural networks with fully parallel architectures is often impractical due to substantial hardware resource demands. This research introduces a hardware-multiplexed, design-parameterised architecture for resource-efficient DNN implementation. Utilizing an enhanced layer-multiplexed architecture, our method allows for realising any neural network size by reusing a single layer. The architecture improves overall system throughput by employing multiplexing to select a necessary configuration. The design, written in Verilog, is implemented on FPGAs and synthesised for ASICs, offering modularity and scalability for various DNN sizes. Control units that handle design parameters based on configuration settings manage the reuse of a single layer. This control unit is critical, as it manages the data flow between layers, from input to intermediate stages, and coordinates the start and finish signals for each layer. It also assigns weights and inputs from previous layers. After processing every layer, the control unit signals the device that the final result is ready for transmission by sending it an "ANN done" signal.

\begin{figure}[!pt]
     \centering
     \includegraphics[scale=0.6]{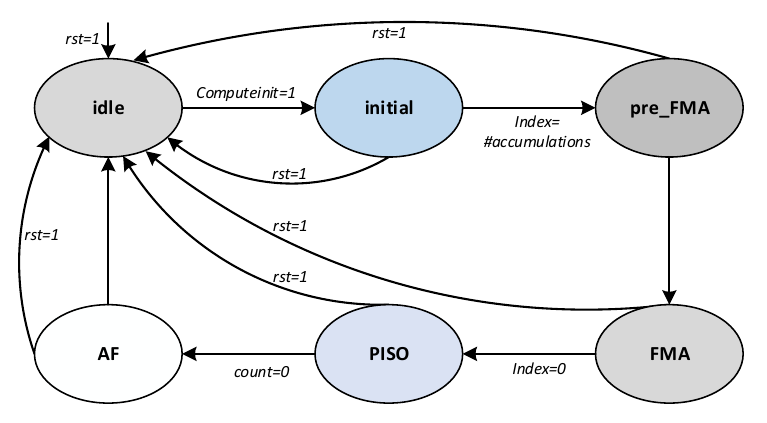}
     \caption{Finite State-machine for the proposed FMA, followed by PISO and the single activation function to be reused with control signals.}
     \label{FMA-fsm}
     
\end{figure}

\subsection{Activation Function Reuse within Layer of Optimised FMA}

Examining a feed-forward DNN with the model number 64:32:32:10 will help you calculate the computation time with the suggested DNN architecture. To address this inefficiency, we have removed the AF from each neuron and connected it to a parallel-in-serial-out (PISO) configuration, which reduces hardware overhead per neuron and the overall area at the expense of some latency. With 64 neurons in the first layer and 196 inputs per neuron, the FMA units need 196 clock cycles to calculate the weighted sum. N FMAs are connected to one reconfigurable AF through PISO in the proposed layer. The proposed AF reuse approach leads to hardware resource savings of (N-1) AF. There is a viable tradeoff between throughput and chip area. The possible challenges may arise in control complexity with the same approach. When applied to a layer-sized network, this method can reduce hardware usage by approximately ten times \cite{ref9,ref7}. While the PISO effectively reduces data throughput, reusing a single AF increases area gain; this small increase in complexity yields significant area savings. 

Typically, the data is fed into input and weight banks using serial multiplexing. After calculating FMAs in parallel, two additional clock cycles are required to process output from the first FMA through the PISO and AF. The first feature map's convolution output is available at the 198\textsuperscript{th} clock cycle. Layer one would take an extra 64 clock cycles to get each output serial and would be stored, thus reducing the bandwidth at the output. The next hidden layer will process the outputs with 32 FMAs for 64 inputs and 66 clock cycles. The layer output would be available after 32 clock cycles in serial manner. Ten clock cycles are needed for the FMA units of the output layer, and additional cycles are needed for the softmax in the output layer, which consists of four neurons with ten inputs. The FMA outputs must be identified before the execution of the layer and once the previous layer has been executed. The approach would help control the engine with easier reconfiguration and power gating for FMAs that are not in use. The mathematical modelling for timing analysis for computation clock cycles required for Fully Parallel architecture (T\textsubscript{P}) and Layer Reuse (T\textsubscript{R}) is shown in equations \ref{TP} and \ref{TR}. Here, L is the number of layers in the DNN model, and n(l) is the number of FMAs in the l\textsuperscript{th} layer.

\begin{equation}
T_P = \sum_{l=1}^{L-1} n(l) + L - 1
\label{TP}
\end{equation}

\begin{equation}
T_R = \sum_{l=1}^{L} n(l) + 2L - 3
\label{TR}
\end{equation}

\begin{table}[!pt]
\caption{FPGA resource utilisation for SOTA architectural works}
\label{arch-resource}
\renewcommand{\arraystretch}{1.5}
\scalebox{0.4}{
\begin{tabular}{cccccccc}
\hline
\textbf{Parameters} & \textbf{Fully Parallel\cite{ref13}} & \textbf{Hardware-Reused\cite{ref12}} & \textbf{AF-Reused\cite{ref7}} & \textbf{Layer-Reused\cite{ref16}} & \textbf{Layer-Multiplexed\cite{ref9}} & \textbf{Proposed} & \textbf{Improvement (\%)} \\\hline
\multicolumn{8}{c}{\textbf{Resource Utilisation}} \\\hline
Slice LUTs & 207908 & 144283 & 208608 & 139561 & 112654 & 13550 & 88 \\
Slice Reg. & 306874 & 155811 & 308374 & 163231 & 113648 & 7962 & 93 \\
BRAM & 53 & 23 & 3 & 24 & 32 & NIL & NIL \\\hline
\multicolumn{8}{c}{\textbf{Network On-Chip   Power (W)}} \\\hline
Logic & 0.914 & 0.256 & 0.843 & 0.238 & 0.133 & 0.01225 & 91.8 \\
Signal & 0.956 & 0.585 & 0.911 & 0.613 & 0.235 & 0.0117 & 95.12 \\
Dynamic & 1.933 & 1.02 & 1.792 & 0.982 & 0.481 & 0.025 & 95.8 \\
Static & 0.261 & 0.252 & 0.58 & 0.251 & 0.248 & 0.242 & 2.24\\\hline
\end{tabular}}
\end{table}

\begin{table}[!pt]
\caption{Quantitative analysis of proposed HYDRA accelerator\\ with prior works}
\label{arch_results_quant}
\renewcommand{\arraystretch}{1.5}
\scalebox{0.55}{
\begin{tabular}{ccccccc}\hline
\textbf{Parameter} & \textbf{Null Hop\cite{ref17}} & \textbf{Mcdanell et.al.\cite{ref18}} & \textbf{Zhu et.al.\cite{ref20}} & \textbf{Lu et.al.\cite{ref21}} & \textbf{RAMAN\cite{ref10}} & \textbf{Proposed} \\\hline
Platform & Xilinx Zynq-7100 & Xilinx VC707 & Xilinx ZCU102 & Xilinx ZC706 & Efinix Ti60 & Xilinx VC707 \\
Model & VGG16 & Custom & ResNet-50 & 1-D CNN & DS-CNN & ANN Layer multiplexed \\
Precision & 16b & N/A & 16b & 16b & 8b & 8b \\\hline
LUTs & 229k & 239k & 390k & 3.24k & 37.2k & 13.5k \\
Registers & 107k & 201k & 278k & N/A & 8.6k & 7.96k \\
DSPs & 128 & 112 & 1352 & 48 & 61 & 0 \\
Power (W) & 1.1 & 2.2 & 15.4 & 0.506 & 0.137 & 0.251\\
Freq. (MHz) & 60 & 170 & 200 & 200 & 75 & 100 \\
GOPS/W & 27.4 & N/A & N/A & 45.05 & 79.68 & 35.21 \\\hline
\end{tabular}}
\end{table}

\section{Experimental Setup \& Performance Evaluation} 

\begin{figure}[!pt]
\centering
\subfloat[]{\includegraphics[width=42.5mm,height=23mm]{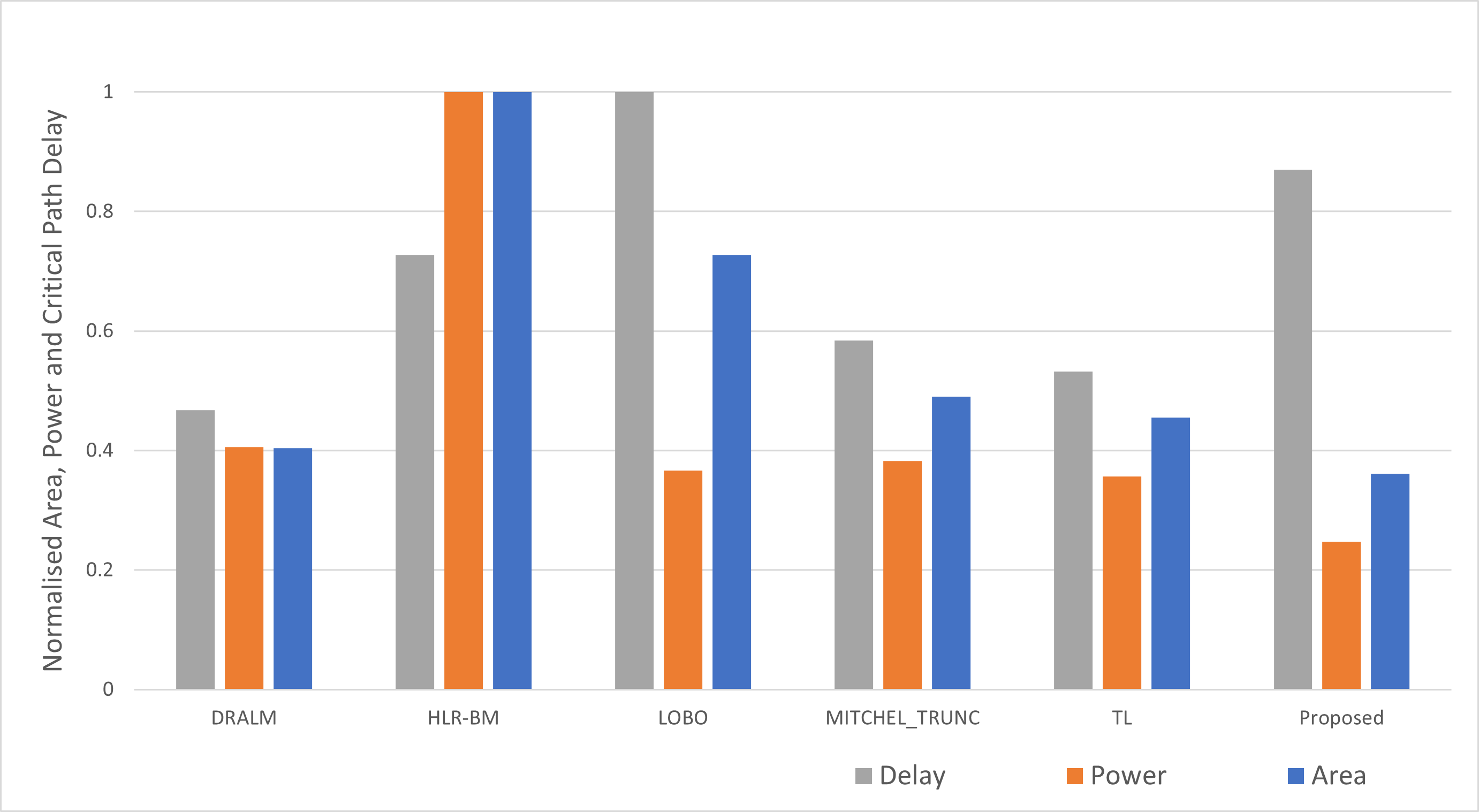}%
\label{PPA-parameters}}
\vspace{-1 mm}
\hfill
\subfloat[]{\includegraphics[width=42.5mm,height=22.5mm]{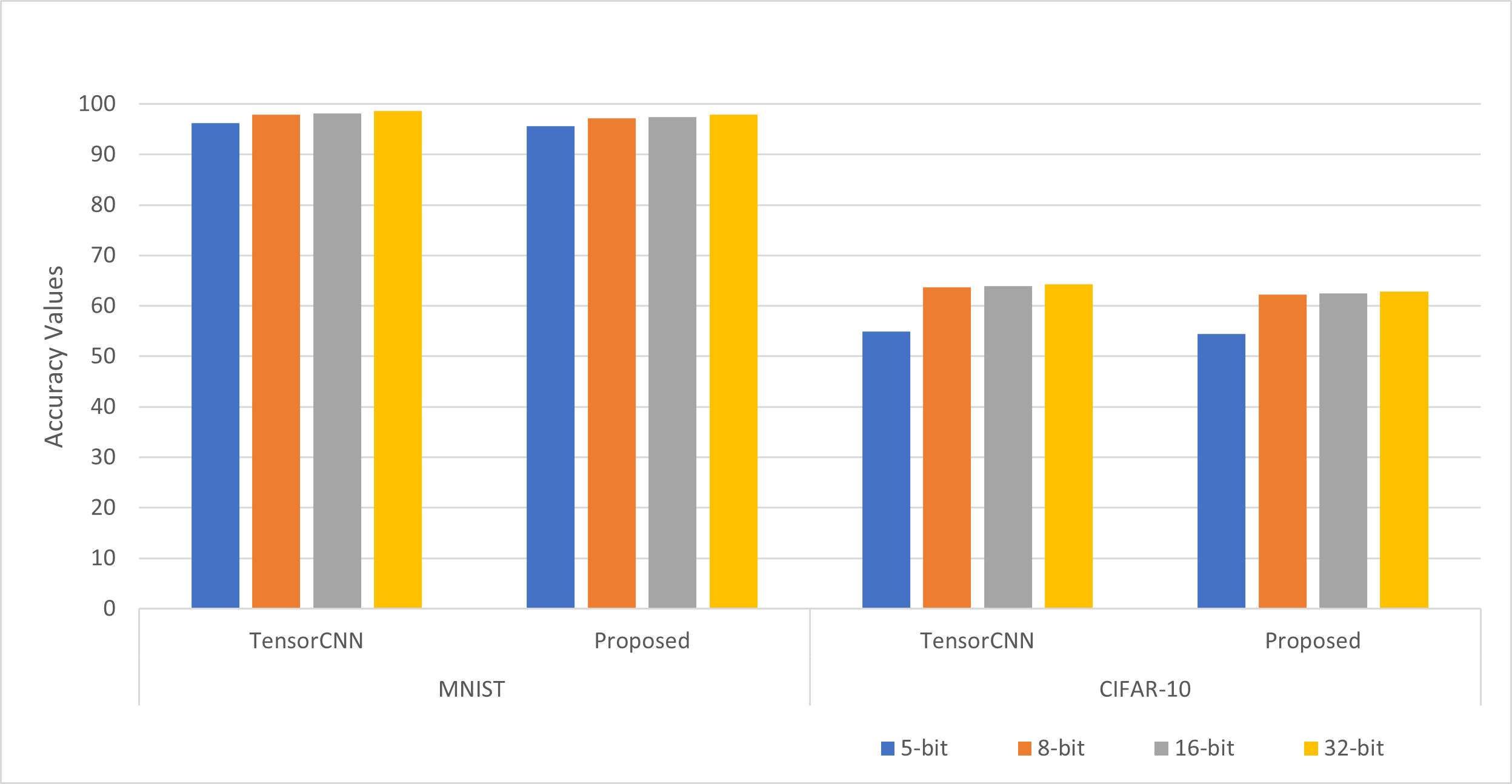}%
\label{accuracy-results}}
\vspace{1 mm}
\caption{ Comparison of (a) Power-Area-Critical Path Delay with State-of-the-Art 8-bit FMA units\cite{ref1, ref2,ref6,ref4,ref5}, (b) Quantisation Impact on Classification Accuracy.}
\vspace{-2 mm}
\end{figure}

Neural networks are widely utilised in everyday applications, and several parameters determine the efficiency of the accelerator. As demonstrated in Table \ref{arch-resource} and Fig. \ref{PPA-parameters}, which offer a comparative analysis with alternative designs, the resource utilization of the FPGA is assessed using the VC707. The resource utilization on the FPGA was evaluated using the VC707, as shown in Table \ref{arch-resource} and Fig. \ref{PPA-parameters}, which provide a comparative analysis with alternative designs. The layer-reuse architecture and activation function (AF) multiplexing, which greatly reduces hardware requirements, are largely responsible for these improvements. Our design demonstrates a reduction in LUTs, slices, and power consumption by 15.4 times, 38.7 times, and 20 times, respectively, compared with the AF-reuse approach \cite{ref15}, which reuses a common AF across fully connected layers. Several other cutting-edge designs, such as Layer multiplexed, AF reused, and hardware-reused architectures \cite{ref9, ref13}, show that LUTs, slices, and power consumption can be reduced by roughly 10 times, 20 times, and 7 times, respectively. 

The additional analysis to discuss the impact of Power-Area-Delay parameters varying with signed bit FMAs is presented in table \ref{arch_results_quant}. For 5, 8, 16, and 32-bit FMAs, the reduction in Slice LUTs are 13\%, 33\%, 24\%, and 15\%, while that of Slice Reg./FF. are by 28\%, 34\%, 31\%, and 30\%, when compared with Xilinx IP. The PDP is improved up to 86 \% at 8-bit precision, with a negligible impact on accuracy, as shown in Table \ref{accuracy-results}. The improved resource efficiency is linearly translated at the architectural level when the proposed custom ANN is applied to MNIST/CIFAR-10. Table \ref{arch-resource} reports the resource comparison with SOTA architectures. Though the SOTA works report resource utilisation for models like VGG-16, MobileNetV2, and ResNet-50, it is also possible in our architecture to implement the same hardware; however, for simplistic implementation, Custom ANN is reported. Our proposed design achieves a noteworthy power efficiency of 35.21 GOPS per watt when operating at 100 MHz. The design highlights the synthesizable nature as no DSP blocks are utilised within. Overall, evaluating the suggested architecture using datasets such as MNIST and CIFAR-10 across various FMAs and bit precisions shows its scalability and flexibility.

\begin{table}[]
\caption{Resource Utilisation for Proposed FMA \\ at Different Quantized Bit-widths }
\label{fpga-util_bit}
\renewcommand{\arraystretch}{1.5}
\scalebox{0.6}{
\begin{tabular}{lcrrrrrrrrr}
\hline

\multirow{2}{*}{\textbf{Bit-width}} & \multicolumn{1}{l}{\multirow{2}{*}{\textbf{FMA}}} & \multicolumn{1}{l}{\multirow{2}{*}{\textbf{\begin{tabular}[c]{@{}l@{}}Slice LUTs\\ (433200)\end{tabular}}}} & \multicolumn{1}{l}{\multirow{2}{*}{\textbf{\begin{tabular}[c]{@{}l@{}}Slice reg.\\ (108300)\end{tabular}}}} & \multicolumn{3}{c}{\textbf{Critical path delay (ns)}} & \multicolumn{3}{c}{\textbf{Power(mW)}} & \multicolumn{1}{l}{\multirow{2}{*}{\textbf{\begin{tabular}[c]{@{}l@{}}PDP\\ (pJ)\end{tabular}}}} \\
 & \multicolumn{1}{l}{} & \multicolumn{1}{l}{} & \multicolumn{1}{l}{} & \multicolumn{1}{l}{\textbf{Delay}} & \multicolumn{1}{l}{\textbf{Logic}} & \multicolumn{1}{l}{\textbf{Signal}} & \multicolumn{1}{l}{\textbf{Logic}} & \multicolumn{1}{l}{\textbf{Signal}} & \multicolumn{1}{l}{\textbf{Dynamic}} & \multicolumn{1}{l}{} \\\hline
\multirow{2}{*}{5-bit} & Xilinx IP & 53 & 28 & 3.090 & 0.813 & 2.277 & 0.27 & 0.21 & 3 & 9.37 \\
 & Proposed & 46 & 20 & 1.259 & 0.378 & 0.881 & 0.18 & 0.24 & 2 & 2.518 \\\hline
\multirow{2}{*}{8-bit} & Xilinx IP & 130 & 44 & 3.816 & 0.921 & 2.895 & 0.66 & 0.60 & 6 & 22.9 \\
 & Proposed & 87 & 29 & 1.307 & 0.363 & 0.945 & 0.48 & 0.54 & 3 & 3.921 \\\hline
\multirow{2}{*}{16-bit} & Xilinx IP & 244 & 60 & 4.451 & 0.966 & 3.485 & 1.26 & 1.26 & 9 & 40.1 \\
 & Proposed & 184 & 41 & 1.509 & 0.274 & 1.235 & 0.96 & 1.16 & 4 & 6.036 \\\hline
\multirow{2}{*}{32-bit} & Xilinx IP & 369 & 76 & 9.051 & 2.268 & 6.783 & 1.95 & 1.95 & 13 & 117.7 \\
 & Proposed & 310 & 53 & 2.632 & 0.323 & 2.309 & 1.86 & 1.86 & 6 & 15.792\\\hline
\end{tabular}}
\end{table}

\section{Conclusion }
This work presents HYDRA architecture, a hybrid data multiplexing and runtime layer-configurable DNN accelerator to enhance the performance within hardware constraints of executing DNNs at edge nodes. By utilizing an innovative layer-multiplexed approach and optimizing the FMA unit, substantial resource consumption is reduced, leading to hardware area reductions of up to 15 times, enhanced power efficiency, and minimal latency. The proposed architectures achieve over 90\% of power consumption and resource utilisation improvements of state-of-the-art works, with 35.21 TOPS\/W. The outcomes highlight how well HYDRA supports resource-constrained edge devices' efficient DNN computations.

\vfill

\end{document}